\RequirePackage{fix-cm}
\documentclass[epj,nopacs]{svjour}   
\smartqed  
\RequirePackage{graphicx}
\RequirePackage{mathptmx}      
\RequirePackage{flushend}
\RequirePackage[numbers,sort&compress]{natbib}
\RequirePackage[colorlinks,citecolor=blue,urlcolor=blue,linkcolor=blue]{hyperref}
\usepackage{revsymb}
\newcommand{\AuAu}    {Au\,+\,Au collisions at 1.23\agev}   
\newcommand{\agev}    {\mbox{$A$~GeV}}               
\newcommand{\gevc}    {\mbox{GeV$/c$}}

\newcommand{\rb}[1]   {\mbox{\textrm{\scriptsize #1}}}
\newcommand{\rbt}[1]  {\mbox{\textrm{\tiny #1}}}
\newcommand{\sqrtsnn} {\ensuremath{\sqrt{s_{_{\rbt{NN}}}}}}

\newcommand{\nch}     {\ensuremath{N_{\rb{ch}}}}
\newcommand{\nhits}   {\ensuremath{N_{\rb{hits}}}}
\newcommand{\nhitsT}  {\ensuremath{N_{\rb{hits}}^{\rbt{TOF}}}}
\newcommand{\nhitsR}  {\ensuremath{N_{\rb{hits}}^{\rbt{RPC}}}}
\newcommand{\nhitsS}  {\ensuremath{N_{\rb{hits}}^{\rbt{TOF+RPC}}}}
\newcommand{\ntracks} {\ensuremath{N_{\rb{tracks}}}}

\newcommand{\npart}   {\ensuremath{N_{\rb{part}}}}
\newcommand{\nspec}   {\ensuremath{N_{\rb{spec}}}}

\newcommand{\npartav} {\ensuremath{\langle N_{\rb{part}} \rangle}}

\newcommand{\bav}     {\ensuremath{\langle b \rangle}}

\newcommand{\sinel}   {\ensuremath{\sigma_{\rb{inel}}}}
\begin{document}
\title{Centrality determination of Au\,+\,Au collisions at 1.23A~GeV with HADES}
%
\titlerunning{Centrality determination with HADES}        
\author{
\vskip 18pt
HADES Collaboration$^{a}$
\vskip 10pt
  J.~Adamczewski-Musch$^{4}$, O.~Arnold$^{10,9}$,
  C.~Behnke$^{8}$, A.~Belounnas$^{15}$, A.~Belyaev$^{7}$,
  J.C.~Berger-Chen$^{10,9}$, J.~Biernat$^{3}$, A.~Blanco$^{2}$,
  C.~~Blume$^{8}$
% ...
}

\institute{
        $^{1}$~Institute of Physics, Slovak Academy of Sciences,
          84228~Bratislava, Slovakia \\
        $^{2}$~LIP-Laborat\'{o}rio de Instrumenta\c{c}\~{a}o e
          F\'{\i}sica Experimental de Part\'{\i}culas,
          3004-516~Coimbra, Portugal \\
        $^{3}$~Smoluchowski Institute of Physics, Jagiellonian
          University of Cracow, 30-059~Krak\'{o}w, Poland \\
        $^{4}$~GSI Helmholtzzentrum f\"{u}r Schwerionenforschung
          GmbH, 64291~Darmstadt, Germany \\
        $^{5}$~Technische Universit\"{a}t Darmstadt,
          64289~Darmstadt, Germany \\
        $^{6}$~Institut f\"{u}r Strahlenphysik, Helmholtz-Zentrum
          Dresden-Rossendorf, 01314~Dresden, Germany \\
        $^{7}$~Joint Institute of Nuclear Research, 141980~Dubna,
          Russia \\
        $^{8}$~Institut f\"{u}r Kernphysik,
          Goethe-Universit\"{a}t, 60438 ~Frankfurt, Germany \\
        $^{9}$~Excellence Cluster "Origin and Structure of the
          Universe", 85748~Garching, Germany \\
        $^{10}$~Physik Department E62, Technische Universit\"{a}t
          M\"{u}nchen, 85748~Garching, Germany \\
        $^{11}$~II.Physikalisches Institut, Justus Liebig
          Universit\"{a}t Giessen, 35392~Giessen, Germany \\
        $^{12}$~Institute for Nuclear Research, Russian Academy
          of Science, 117312~Moscow, Russia \\
        $^{13}$~Institute of Theoretical and Experimental
          Physics, 117218~Moscow, Russia \\
        $^{14}$~Department of Physics, University of Cyprus,
          1678~Nicosia, Cyprus \\
        $^{15}$~Institut de Physique Nucl\'{e}aire, CNRS-IN2P3,
          Univ. Paris-Sud, Universit\'{e} Paris-Saclay, F-91406~Orsay
          Cedex, France \\
        $^{16}$~Nuclear Physics Institute, The Czech Academy of
          Sciences, 25068~Rez, Czech Republic \\
        $^{17}$~LabCAF. F. F\'{\i}sica, Univ. de Santiago de
          Compostela, 15706~Santiago de Compostela, Spain \\
	\\
        $^{a}$~e-mail: hades-info@gsi.de \\
        $^{b}$~also at ISEC Coimbra, ~Coimbra, Portugal \\
        $^{c}$~also at Technische Universit\"{a}t Dresden,
          01062~Dresden, Germany \\
        $^{d}$~also at ExtreMe Matter Institute EMMI,
          64291~Darmstadt, Germany \\
        $^{e}$~also at Utrecht University, 3584 CC~Utrecht, The
          Netherlands \\
        $^{f}$~also at Moscow Engineering Physics Institute
          (State University), 115409~Moscow, Russia \\
        $^{g}$~also at Frederick University, 1036~Nicosia,
          Cyprus \\
        $^{h}$~also at Dipartimento di Fisica and INFN,
          Universit\`{a} di Torino, 10125~Torino, Italy \\
        \\
         $^{\dag}$ Deceased.\\
        \\}

\authorrunning{J.~Adamczewski-Musch et al.} 
%
%
\date{Received: 23 January 2018 / Revised: 10 April 2018}
%
%
\abstract{
The centrality determination for \AuAu, as measured with HADES at the
GSI-SIS18, is described. In order to extract collision geometry
related quantities, such as the average impact parameter or number of
participating nucleons, a Glauber Monte Carlo approach is employed.
For the application of this model to collisions at this relatively low
centre-of-mass energy of $\sqrtsnn = 2.42$ GeV special investigations
were performed.  As a result a well defined procedure to determine
centrality classes for ongoing analyses of heavy-ion data is
established.
%
} 
\maketitle
%

%
\section{Introduction}
\label{intro}

Heavy-ion collisions at relativistic and ultra-relativistic energies
allow to study highly excited, strongly interacting matter.  Due to
the extended volume of the colliding nuclei, the size of the
interaction volume varies strongly with impact parameter $b$, defined
as the distance between the centers of the nuclei in a plane
transverse to the beam axis. Therefore, it is crucial to define a
framework which allows to relate observable quantities, such as the
measured charge particle multiplicity, to the centrality of the
collision.  The centrality $C$ is defined as the fraction of the total
cross section $\sigma_{AA}$ and is directly related to the impact
parameter 
\begin{equation}
    C = \frac{\int_{0}^{b}     d \sigma/d b^{\prime} \: d b^{\prime}}
         {\int_{0}^{\infty} d \sigma/d b^{\prime} \: d b^{\prime}}
      = \frac{1}{\sigma_{AA}} 
        \int_{0}^{b} \frac{d \sigma}{d b^{\prime}} \: d b^{\prime} , 
\end{equation}
with the differential cross section $\frac{d \sigma}{d b}$.  On one
side this requires a good determination of $\sigma_{AA}$ and on the
other side one has to establish a relation between $b$ and the
observable multiplicity $N$, such that the latter can be sorted
according to the corresponding fraction of the total cross section
\begin{equation}
    C \approx \frac{1}{\sigma_{AA}} 
        \int_{N^{\rb{thr}}}^{\infty} \frac{d \sigma}{d N^{\prime}} \: d N^{\prime} , 
\end{equation}
where $N^{\rb{thr}}$ is the lower multiplicity threshold of a given class.
Experiments at high energies usually employ the Glauber model
\cite{Glauber:1955qq,Glauber:1959,Glauber:2006gd} for this purpose, as
described, e.g., in refs.~\cite{Miller:2007ri,Abelev:2013qoq}.  In this
approach, heavy-ion collisions are treated as a superposition of
independent nucleon-nucleon interactions.  Following the eikonal
approximation, the trajectories of the individual nucleons are assumed
to be straight lines, and a nucleon is defined as a participant, if it
experiences at least one binary collision along its path.  The number
of participating nucleons \npart\ can thus be used to quantify the
size of the interaction volume.  The number of spectators, i.e. of
non-interacting nucleons, is related to the participants by $\nspec =
A_{\rb{proj}} + A_{\rb{targ}} - \npart$, where $A_{\rb{proj}}$
($A_{\rb{targ}}$) is the nucleon number of the projectile (target)
nucleus.  Following the assumptions of the wounded nucleon model
\cite{Bialas:1976ed}, the number of produced particles should be
directly proportional to \npart.

%
\begin{figure}
\begin{center}
\includegraphics[width=1.\linewidth]{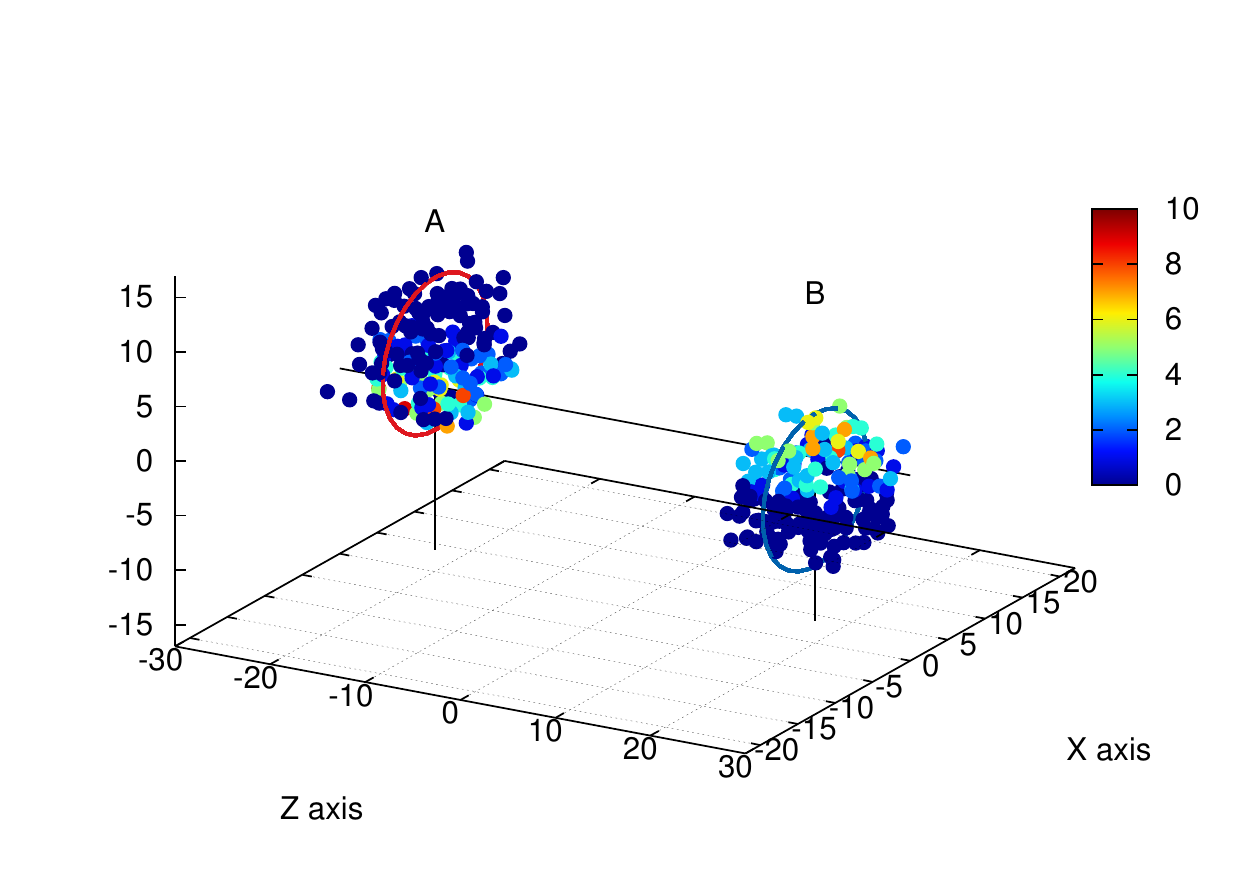}
\end{center}
\caption{Spatial distribution of the nucleons before the collision of
  nuclei A and B as generated with the Glauber MC model with an impact
  parameter $b = 6$~fm.  The beam direction is along the $Z$-axis.
  The color scheme encodes the number of inelastic collisions that a
  single nucleon experiences in this particular collision process.  The
  size of the nucleons corresponds to the inelastic nucleon-nucleon
  cross section and the radii of the circles to the one of the
  gold nuclei.}
\label{fig:glaubermc}
\end{figure}
%

This relation thus allows to fit measured multiplicity distributions
with a Glauber Monte Carlo (MC) model, in which many nucleus-nucleus
collisions are statistically sampled.  Based on this fit, centrality
classes with their corresponding average impact parameter \bav\ and
average number of participants \npartav\ can be derived by selecting
intervals of event multiplicity.  The aim of the procedure described
in this article is to provide a framework which allows to categorize
events measured by HADES according to their multiplicity in centrality
classes and to relate these classes to \bav\ and \npartav.

While the validity of the Glauber model at higher energies
(i.e. \sqrtsnn\ around 10~GeV and above) is well established, the
applicability at lower energies is not immediately evident.  The
semi-classical description of the eikonal approximation requires that
the reduced de~Broglie wavelength, $\lambdabar_{B} = \hbar/p$, where
$p$ is the characteristic momentum of the nucleon, is small relative
to the size of the nucleus.  This condition is clearly satisfied for a
nucleon of $\sim 2$~\gevc\ momentum traversing a gold nucleus at rest,
a situation close to Au\,+\,Au collisions at 1.23\agev, where
\begin{equation}
  \lambdabar_{B} \approx 0.1 \:\,\textrm{fm} 
  \: \ll \:
  R_{Au} \approx 5.4 \:\,\textrm{fm}.
\end{equation}
Indeed, it was claimed that, with appropriate corrections of the order
of 15\%, the validity of the eikonal Glauber model can even be
extended down to energies as low as about 45 MeV \cite{Buuck:2014era}.

In this article, we describe the application of a Glauber MC model to
Au\,+\,Au data at 1.23\agev, as measured with HADES, the High-Acceptance
Di-Electron Spectrometer, which is installed at the
SIS18\footnote{Schwer-Ionen-Synchrotron.} accelerator at the
GSI\footnote{Gesellschaft f\"{u}r Schwerionenforschung.} in Darmstadt,
Germany.  The implementation and the adjustments to the model,
necessary for this data set, are discussed in
sect.~\ref{sect:glauber-model} and the experimental observables for
the charged particle multiplicity as used for the centrality
determination are presented in sect.~\ref{sect:experimental}. In
sect.~\ref{sect:cent_deter} we present the method and in
sect.~\ref{sect:syst_error} the systematic uncertainties on the
centrality determination.  The corresponding centrality classes are
shown in sect.~\ref{sect:cent-class} and a conclusion is drawn in
sect.~\ref{sect:conclution}.

%
%
%
\section{Glauber Monte Carlo model implementation}
\label{sect:glauber-model}

The analysis discussed in the following is based on the implementation
of a Glauber MC model as described in
refs.~\cite{Miller:2007ri,Loizides:2014vua}.  In this approach, the
following steps are processed event-by-event:

\begin{enumerate}
\item The impact parameter $b$ of a given collision is selected
  randomly according to a probability distribution $P(b)\propto b\:db$
   in the range from 0 to $b_{\rb{max}}$, where $b_{\rb{max}}$ has
  to be larger than the sum of the radii of the projectile and target
  nucleus, $b_{\rb{max}} \ge R_{\rb{proj}} + R_{\rb{targ}}$.
\item For each of the nuclei, the $N_{\rb{proj}}$ and $N_{\rb{targ}}$
  nucleons are positioned randomly, within spheres of radii
  $R_{\rb{proj}}$ and $R_{\rb{targ}}$.  This is achieved by using a
  uniform probability distribution in the azimuthal ($\phi$) and polar
  ($\cos \theta$) angles, and a radial density distribution given by
  $P(r) \propto r^{2} \: \rho(r)$, where $\rho(r)$ is specified in
  eq.~\ref{eq:fermi-dist}.
\item The collision process itself is evaluated such that binary
  combinations of all nucleons in the two nuclei are formed and a
  decision is made whether a collision is actually taking place
  between them.  This decision is based on the criterion that the
  distance between nucleon $i$ and $j$ in the transverse plane,
  \begin{equation}
    d_{ij} = \sqrt{\Delta x^{2} + \Delta y^{2}} ,
  \end{equation}
  is smaller than the
  radius defined by the inelastic nucleon-nucleon cross section (black
  disk approximation):
  \begin{equation}
    d_{ij}^{2} \le \sinel^{\rb{NN}} / \pi .
  \end{equation}
\end{enumerate}

Figure~\ref{fig:glaubermc} shows an example for the spatial
configuration of two colliding nuclei obtained with the above
described procedure.  In order to apply the Glauber MC model to the
relatively low centre-of-mass energies of the heavy-ion collisions
under consideration here, several adjustments had to be performed
\cite{Kardan:2015}.  To parametrize the radial charge density
distributions a two-parameter Fermi distribution is used:
\begin{equation}
\label{eq:fermi-dist}
   \rho(r) = \frac{1 + w \: (r / R)^{2}}
                  {1 + \exp \left( \frac{r - R}{a} \right)} .
\end{equation}
For the gold nuclei, we use the parameters $R = 6.55$~fm and $a =
0.52$~fm, which best describe the measurements in \cite{LB:2004}.
The parameter $w$, which is used to describe nuclei whose density is
lower at the centre than in the outer regions, is here set to $w =
0$.  To make sure that two nucleons cannot overlap in space, a new
position can be reassigned to one of them during step 2. of the above
described MC procedure, if their distance falls below a minimal
separation distance $d_{\rb{min}}$.  Here we use a value of
$d_{\rb{min}} = 0.9$~fm, as it facilitates a consistent description of
the total cross section.  However, as pointed out in
\cite{Alvioli:2011sk}, $d_{\rb{min}} = 0$ might be more suitable for a
proper calculation of eccentricities.  While at high energies the
inelastic nucleon-nucleon cross section is only weakly dependent on
\sqrtsnn, it exhibits a rapid variation with energy in the energy
regime discussed here.  Possible variations thus constitute a
significant contribution to the systematic uncertainty (see
sect.~\ref{sect:syst_error}).  Based on the data collected in
\cite{Agashe:2014kda} and parametrized in \cite{Bystricky:1987} we use
the following values as default: $\sinel^{\rb{pp}} = 26.4$~mb and
$\sinel^{\rb{np}} = 21.0$~mb.  Assuming isospin symmetry
(i.e. $\sinel^{\rb{pp}} = \sinel^{\rb{nn}}$ and $\sinel^{\rb{np}} =
\sinel^{\rb{pn}}$), we construct the isospin averaged nucleon-nucleon
cross section for a gold nucleus as $\sinel^{\rb{NN}} = 23.8$~mb.


%
\section{Experimental observables}
\label{sect:experimental}
The setup of the HADES experiment is described in detail in
\cite{Agakishiev:2009am}.  HADES is a charged particle detector
consisting of a six-coil toroidal magnet centered around the beam axis
and six identical detection sections located between the coils and
covering polar angles between $18^{\circ}$ and $85^{\circ}$.  Each
sector is equipped with a Ring-Imaging Cherenkov (RICH) detector
followed by four layers of Multi-Wire Drift Chambers (MDCs), two in
front of and two behind the magnetic field, as well as a scintillator
Time-Of-Flight detector (TOF) ($45^{\circ}$~--~$85^{\circ}$) and
Resistive Plate Chambers (RPC) ($18^{\circ}$~--~$45^{\circ}$).  Hadron
identification is based on the time-of-flight measured with TOF and
RPC, and on the energy-loss information from TOF as well as from the
MDC tracking chambers.  Electron candidates are in addition selected
via their signals in the RICH detector.  Combining these information
with the track momenta, as determined from the deflection of the
tracks in the magnetic field, allows to identify charged particles
(e.g. pions, kaons or protons) with a high significance.

Several triggers are implemented to start the data acquisition.  The
minimum bias trigger is defined by a signal in the START detector
in the beam line (CVD\footnote{Chemical Vapor Deposition.} diamond
detector).  In addition, online Physics Triggers (PT) are used, which
are based on hardware thresholds on the TOF signals, proportional to
the event multiplicity, corresponding to at least 5 (PT2) or 20 (PT3)
hits in the TOF.  Events are selected offline by requiring that their
global event vertex is inside the target region, i.e. between $z =
-65$~mm and $0$~mm along the beam axis.

For the centrality determination, currently two different experimental
observables are considered within HADES, both of which provide a
measurement of the charged particle density close to mid-rapidity.
One is the number of tracks reconstructed with the MDCs, \ntracks,
while the other is based on the summed number of hits detected by the
TOF and the RPC detectors, $\nhitsS = \nhitsT + \nhitsR$.  The first
one has the advantage of being less contaminated by background hits, but
requires a full reconstruction of the tracks in all MDCs.  \ntracks\
contains only track candidates which do not share any space points
with other tracks, have a good matching of track to point position
and have a distance-of-approach relative to the global event vertex of
less than 10~mm.  These cuts provide a very clean track sample but
also significantly reduce the available multiplicity.  The second one,
\nhits, provides generally a larger and more stable acceptance and
thus a better statistical significance than \ntracks.  Therefore it is
the default observable for centrality selection in most analyses of
HADES data.  However, in order to assure that the larger background
from secondary hits included in \nhits\ does not distort the
centrality determination, we investigate in the following both
observables in comparison.

%
\begin{figure}
\begin{center}
\includegraphics[width=0.9\linewidth]{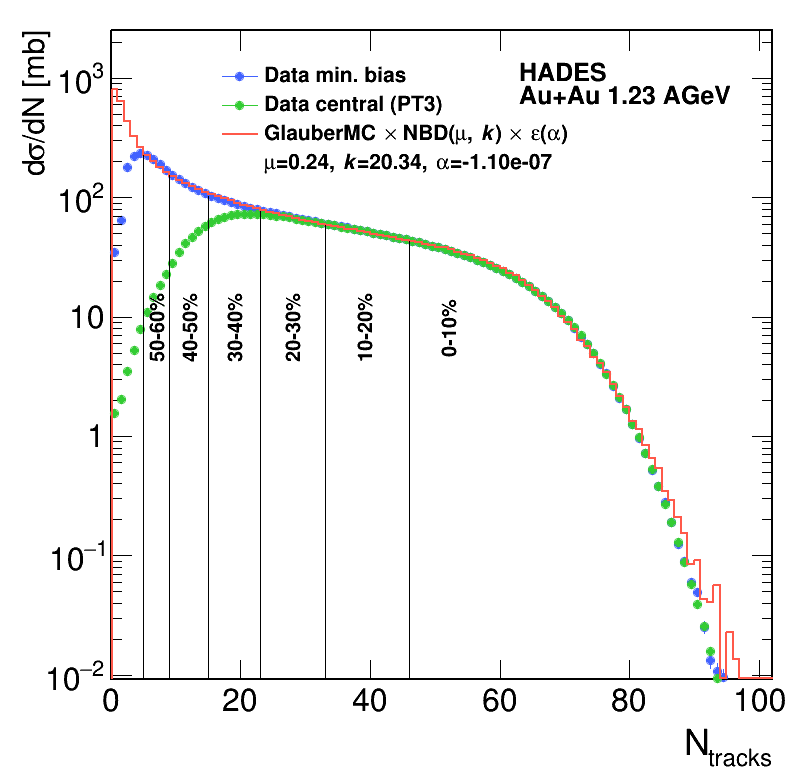}
\includegraphics[width=0.9\linewidth]{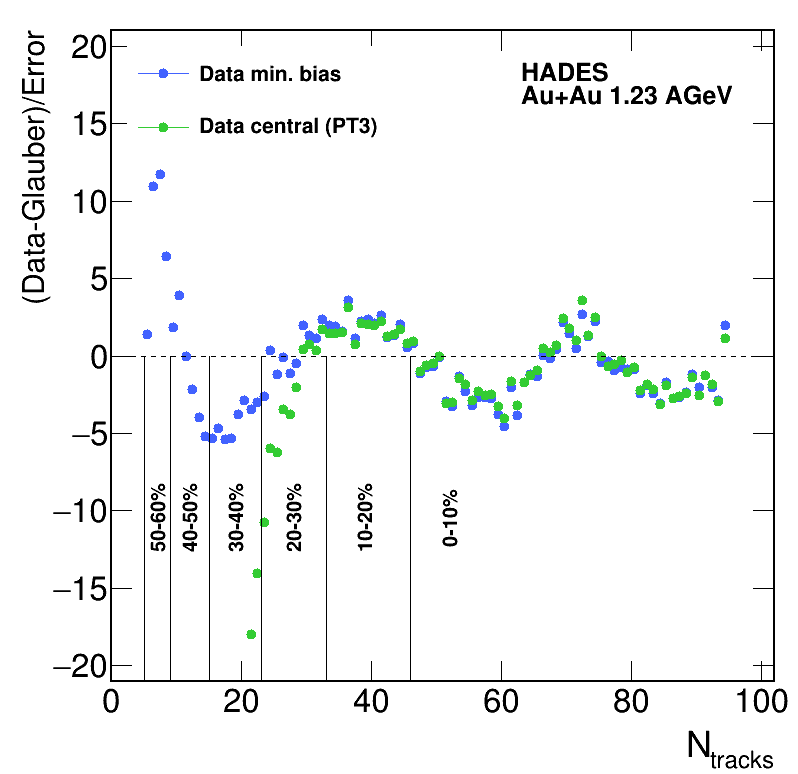}
\end{center}
\caption{Upper panel: The cross section as a function of \ntracks\ for
  minimum bias (blue symbols) and central (PT3 trigger, green symbols)
  data in comparison with a fit using the Glauber MC model (red
  histogram).  The centrality classes are represented by the vertical
  lines.  See text and table~\ref{tab:fit_parameters} for a
  description of the parameters $\mu$, $k$ and $\alpha$.
  Lower panel: The difference between data and Glauber MC model
  divided by the quadratic sum of both errors.}
\label{fig:ntracks_data}
\end{figure}
%

%
\begin{figure}
\begin{center}
\includegraphics[width=0.9\linewidth]{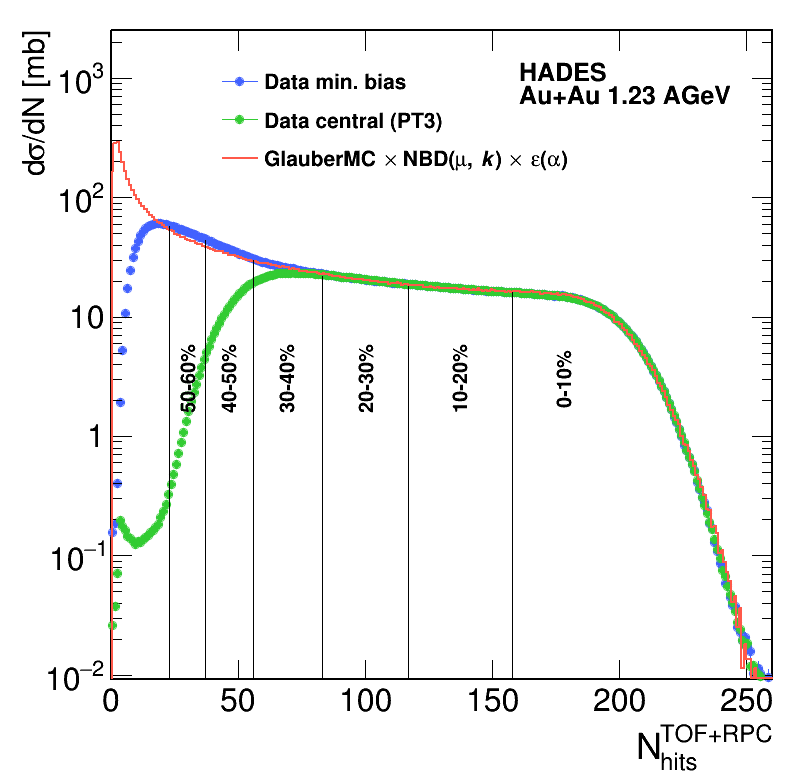}
\includegraphics[width=0.9\linewidth]{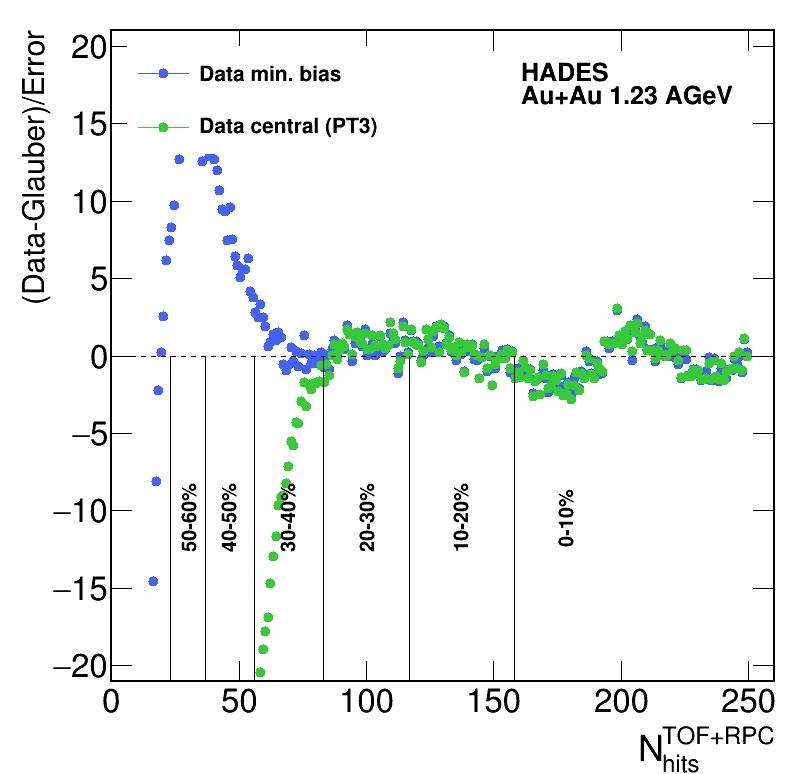}
\end{center}
\caption{Upper panel: The cross section as a function of \nhitsS\ for
  the sum of TOF and RPC hits.  The upper panel shows minimum bias
  (blue symbols) and central (PT3 trigger, green symbols) data in
  comparison with the Glauber MC model (red histogram).  See text and
  table~\ref{tab:fit_parameters} for a description of the parameters
  $\mu$, $k$ and $\alpha$.
  Lower panel: The difference between data and Glauber MC model
  divided by the quadratic sum of both errors.}
\label{fig:nhits_data_total}
\end{figure}
%

%
\begin{figure}
\begin{center}
\includegraphics[width=0.9\linewidth]{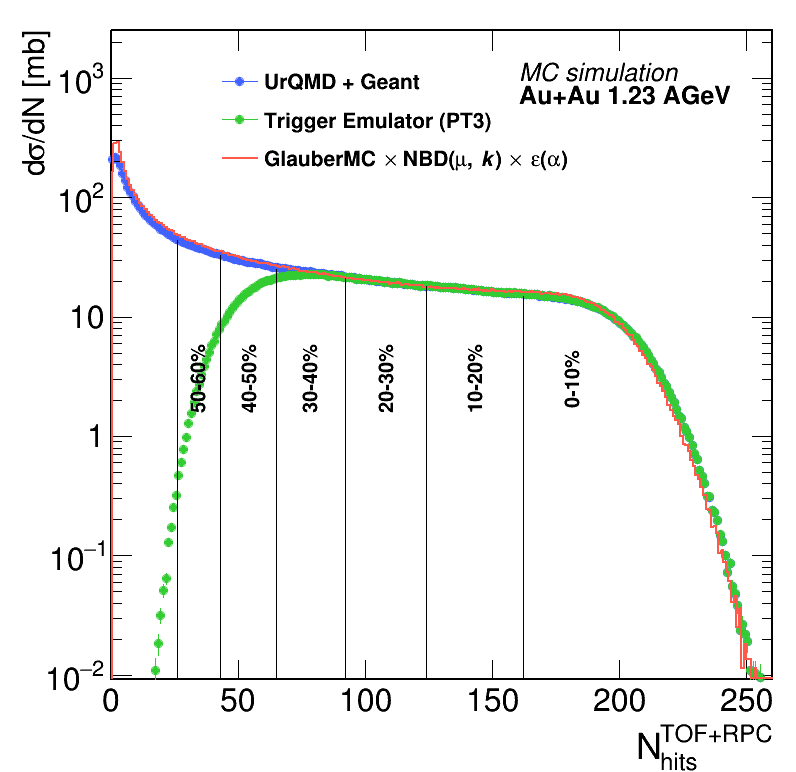}
\end{center}
\caption{The cross section as a function of \nhitsS\ for the sum of
  TOF and RPC hits.  The panel shows a simulation based on UrQMD events
  filtered through the detailed detector simulation (blue symbols) and
  in addition through an emulator of the PT3 trigger (green symbols) 
  in comparison with the Glauber MC model (red histogram) also show in
  fig.\ref{fig:nhits_data_total}.  See text and
  table~\ref{tab:fit_parameters} for a description of the parameters
  $\mu$, $k$ and $\alpha$.}
\label{fig:nhits_sim_total}
\end{figure}
%
%

%
\begin{figure}
\begin{center}
\includegraphics[width=0.9\linewidth]{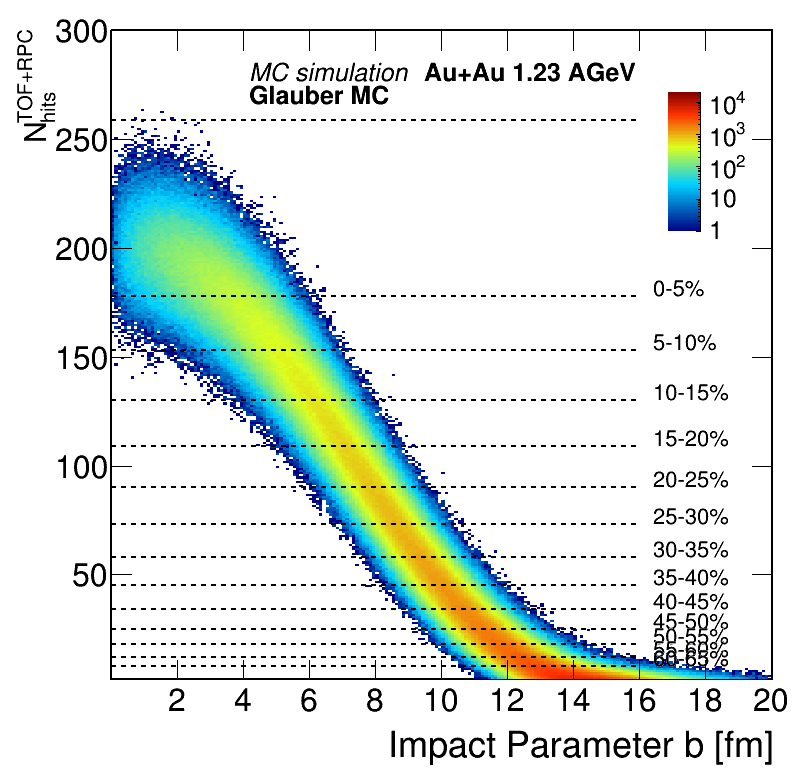}
\includegraphics[width=0.9\linewidth]{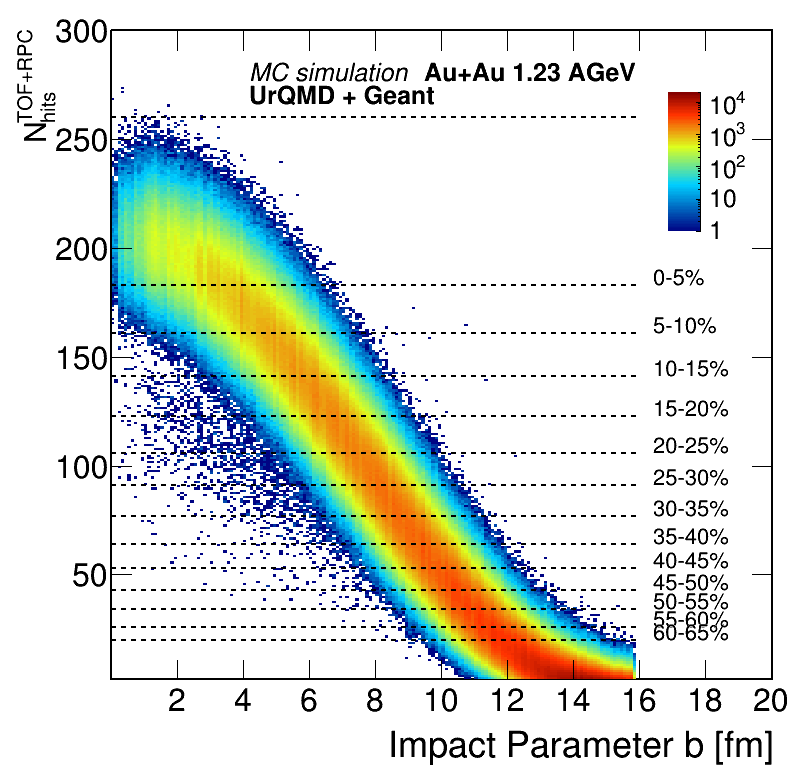}
\end{center}
\caption{The anti-correlation between the total number of hits \nhitsS\
  in TOF and RPC and the impact parameter $b$.  The different
  centrality classes are indicated by the dashed lines.  The upper
  panel is calculated with the Glauber MC model while the lower one is
  obtained from simulated UrQMD events filtered through the detector
  simulation.}
\label{fig:nhits_correlation_urqmd}
\end{figure}
%

%

%
\section{Centrality determination}
\label{sect:cent_deter}
Following the assumption of the wounded nucleon
model~\cite{Bialas:1976ed}, the measured charged particle
multiplicity, \nch, should on average be directly proportional to
\npart.  To allow for event-by-event fluctuations around this average
value, the number of charged particles per participant is sampled for
each MC participant from a Negative Binomial probability Distribution
(NBD) with a mean $\mu$, i.e.
\begin{equation}
  P_{\mu,k}(n) =  \frac{\Gamma(n+k)}{\Gamma(n+1)\Gamma(k)} 
                \cdot
                \frac{(\mu/k)^n}{(\mu/k+1)^{n+k}}\,,
\end{equation}
where $\Gamma$ is the gamma function and the dispersion parameter
$k$ is related to the relative width by $\sigma / \mu = \sqrt{1/\mu +
  1/k}$.  After summing over all participants in a given event this
procedure yields on average the proportionality
$\langle \nch \rangle = \mu \cdot \npartav$.  The parameters $\mu$ and
$k$ take into account the acceptance, reconstruction efficiency and
resolution of the specific observables ($\nch = \ntracks$ or \nhits)
and are determined by a minimization procedure in which the simulated
\nch\ distribution is fitted to the measured one.  

In order to take additional, non-linear multiplicity dependent
inefficiencies into account, the value sampled from the NBD is further
folded with a phenomenological efficiency function $\epsilon(\alpha) =
1 - \alpha \cdot \npart^{2}$.  This function models the relative
variation of the efficiency for charged tracks and parametrizes the
corresponding efficiency obtained from events simulated with the
transport model UrQMD \cite{Bass:1998ca} filtered through a detailed
simulation of the detector response based on GEANT3.21
\cite{Brun:1994aa}.  In case of \nhits\ it also takes into account the
additional contribution from secondary particles produced in the
detector material.  Even for very central events $\epsilon$ differs
not more than 20\% from unity.  Table~\ref{tab:fit_parameters}
summarizes the parameters used for the different observables.  Please
note that, for \nhitsS, the individual numbers \nhitsT\ and \nhitsR\
had to be fit separately, since the two measurements are subject to
different systematic effects.

\begin{table}[ht]
\renewcommand{\arraystretch}{1.3} 
\centering
\begin{tabular}{lccc}
\hline\hline
Observable & \multicolumn{3}{c}{Parameter}       \\
           & $\mu$ & $k$   & $\alpha$            \\ 
\hline
\ntracks   & 0.24  & 20.34 & $1.10 \cdot 10^{-7}$ \\
\nhitsT    & 0.20  &  6.36 & $1.64 \cdot 10^{-6}$ \\
\nhitsR    & 0.50  & 29.06 & $1.64 \cdot 10^{-6}$ \\
\hline
\end{tabular}
\caption{The fit parameters obtained for the different observables.}
\label{tab:fit_parameters}
\end{table}

%
\begin{figure}
\begin{center}
\includegraphics[width=0.9\linewidth]{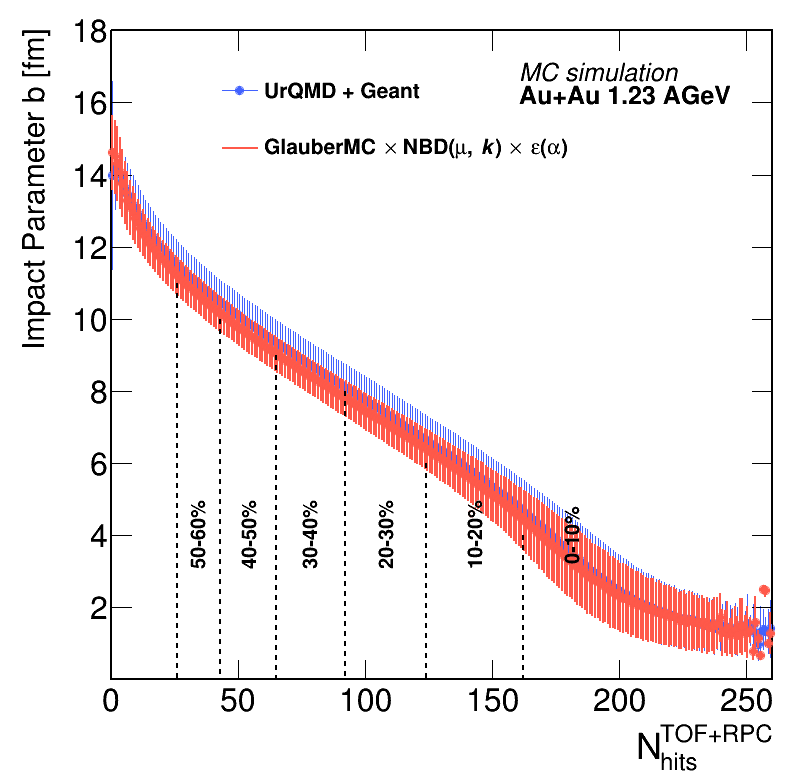}
\includegraphics[width=0.9\linewidth]{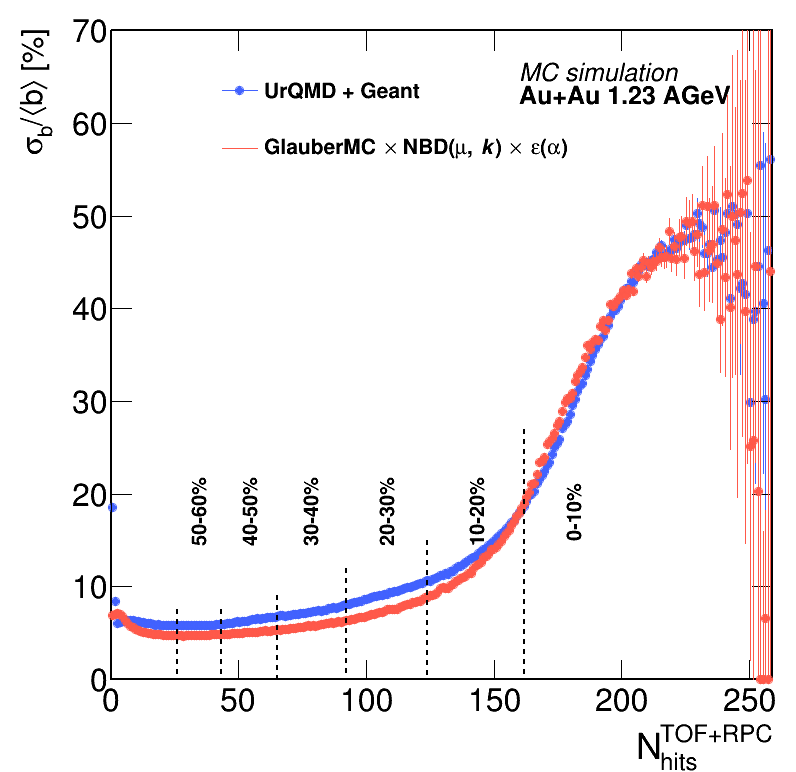}
\end{center}
\caption{Upper panel: The correlations between the average impact 
  parameter $\langle b \rangle$ and the total number of hits \nhitsS\
  in TOF and RPC, as obtained from simulated UrQMD events filtered 
  through the detector simulation.  The error bars shown here are the
  Gaussian $\sigma$.  The different centrality classes are indicated
  by the dashed lines.  In one case, $b$ is taken directly from UrQMD
  (blue symbols), while in the other case (red symbols) it is derived
  from the Glauber MC model fit.
  Lower panel: The corresponding resolution in $b$, defined as the
  ratio of the Gaussian $\sigma$ and the mean of the impact parameter
  distributions \bav.}
\label{fig:nhits_resolution}
\end{figure}
%

The upper panel of fig.~\ref{fig:ntracks_data} shows the comparison of
the fit result to the measured \ntracks\ distribution.  As illustrated
in the lower panel of the figure, a good description is achieved for
the 60\% most central part of the minimum bias data.  Also, the
largest part (i.e. most central 30\%) of the PT3~triggered events is
fitted quite well.  Below these centralities the corresponding trigger
thresholds set in, which are not included in the presented Glauber MC
model.  The PT3~trigger covers the most central $43$\% of the total
cross section, whereas the PT2~trigger includes $72$\%.  Most physics
analyses are based on PT3~triggered data, which are further restricted
to the $0 - 40$\% most central event class or smaller subclasses.

For the measured \nhitsS\ distribution the comparison to the Glauber
MC model is shown in the upper panel of
fig.~\ref{fig:nhits_data_total}.  Also here a very good agreement
is achieved in the same regions of the cross section, as demonstrated
in the lower panel.  The fit has also been performed on UrQMD events,
processed by the full detector simulation and reconstruction chain, as
well as an additional emulator of the PT3~trigger (green symbols).  As
shown in fig.~\ref{fig:nhits_sim_total} the simulated data can be
well described by the same Glauber MC model fit, illustrating that our
centrality selection procedure provides a good correspondence between
event classes defined in data and in models.

%
\begin{table*}[t]
\begin{tabular*}{\textwidth}{@{\extracolsep{\fill}}rrrcrcccc@{}}
\hline\hline
\multicolumn{1}{l}{Centrality}                &
\multicolumn{1}{l}{$b_{\rb{min}}$}               &
\multicolumn{1}{l}{$b_{\rb{max}}$}               &
\multicolumn{1}{l}{\bav}                      &
\multicolumn{1}{l}{\npartav}                  &
\multicolumn{1}{l}{RMS(\npart)}               &
\multicolumn{3}{c}{Uncertainties on \npartav} \\
\multicolumn{1}{l}{Classes}                   &
\multicolumn{1}{c}{ }                         &
\multicolumn{1}{c}{ }                         &
\multicolumn{1}{c}{ }                         &
\multicolumn{1}{c}{ }                         &
\multicolumn{1}{c}{ }                         &
\multicolumn{1}{c}{Model}                     &
\multicolumn{1}{c}{\nhitsS}                   &
\multicolumn{1}{c}{\ntracks}                  \\
\hline\hline 
  0 --  5~\%  &  0.00 &  3.30 &  2.20 & 331.3 & 19.4 & $\pm$  10.4 & $\pm$ 10.6 & $\pm$ 19.1 \\
  5 -- 10~\% &  3.30 &  4.70 &  4.04 & 275.6 & 16.4 & $\pm$  11.4 & $\pm$ 11.1 & $\pm$  7.6 \\
 10 -- 15~\% &  4.70 &  5.70 &  5.22 & 231.9 & 13.7 & $\pm$   9.2 & $\pm$ 10.4 & $\pm$  9.2 \\
 15 -- 20~\% &  5.70 &  6.60 &  6.16 & 195.5 & 13.0 & $\pm$   7.6 & $\pm$  9.5 & $\pm$  7.4 \\
 20 -- 25~\% &  6.60 &  7.40 &  7.01 & 163.3 & 12.2 & $\pm$   7.5 & $\pm$  8.4 & $\pm$  7.8 \\
 25 -- 30~\% &  7.40 &  8.10 &  7.75 & 135.8 & 11.4 & $\pm$   7.7 & $\pm$  8.0 & $\pm$  7.9 \\
 30 -- 35~\% &  8.10 &  8.70 &  8.40 & 113.2 & 10.6 & $\pm$   6.1 & $\pm$  6.6 & $\pm$  5.7 \\
 35 -- 40~\% &  8.70 &  9.30 &  9.00 &  93.7 & 10.5 & $\pm$   4.5 & $\pm$  5.3 & $\pm$  4.8 \\
 40 -- 45~\% &  9.30 &  9.90 &  9.60 &  75.5 & 10.1 & $\pm$   4.8 & $\pm$  5.8 & $\pm$  4.5 \\
 45 -- 50~\% &  9.90 & 10.40 & 10.15 &  60.4 &  9.4 & $\pm$   4.8 & $\pm$  5.3 & $\pm$  3.6 \\
 50 -- 55~\% & 10.40 & 10.90 & 10.65 &  48.0 &  8.9 & $\pm$   3.6 & $\pm$  4.3 & $\pm$  2.8 \\
 55 -- 60~\% & 10.90 & 11.40 & 11.15 &  36.9 &  8.3 & $\pm$   2.5 & $\pm$  3.7 & $\pm$  3.4 \\
\hline\hline 
  0 -- 10~\% &  0.00 &  4.70 &  3.14 & 303.0 & 33.1 & $\pm$ 11.0 & $\pm$ 12.0 & $\pm$ 15.3 \\
 10 -- 20~\% &  4.70 &  6.60 &  5.70 & 213.1 & 22.6 & $\pm$ 11.1 & $\pm$ 11.5 & $\pm$ 11.5 \\
 20 -- 30~\% &  6.60 &  8.10 &  7.38 & 149.8 & 18.1 & $\pm$  9.7 & $\pm$ 10.0 & $\pm$ 10.0 \\
 30 -- 40~\% &  8.10 &  9.30 &  8.71 & 103.1 & 14.4 & $\pm$  6.8 & $\pm$  7.5 & $\pm$  8.2 \\
 40 -- 50~\% &  9.30 & 10.40 &  9.86 &  68.4 & 12.3 & $\pm$  6.5 & $\pm$  7.0 & $\pm$  7.7 \\
 50 -- 60~\% & 10.40 & 11.40 & 10.91 &  42.3 & 10.2 & $\pm$  4.7 & $\pm$  4.2 & $\pm$  4.5 \\
\hline\hline 
  0 -- 20~\% &  0.00 &  6.60 &  4.40 & 258.8 & 53.2 & $\pm$ 11.0 & $\pm$  9.0 & $\pm$ 11.6 \\
 20 -- 40~\% &  6.60 &  9.30 &  8.03 & 127.1 & 28.5 & $\pm$  7.1 & $\pm$  7.1 & $\pm$  7.8 \\
 40 -- 60~\% &  9.30 & 11.40 & 10.39 &  55.3 & 17.3 & $\pm$  5.1 & $\pm$  5.1 & $\pm$  3.8 \\
\hline\hline 
  0 -- 40~\% &  0.00 &  9.30 &  6.20 & 193.3 & 78.5 & $\pm$  8.0 & $\pm$  7.7 & $\pm$  7.8 \\
\hline\hline 
\end{tabular*}
\caption{Centrality classes in fixed intervals of impact parameter
  $b_{\rb{min}} - b_{\rb{max}}$ for \AuAu.
  Listed are the mean impact parameter $\langle b\rangle$, mean number
  of participants \npartav\ and the RMS of the \npart\ distributions
  in the different centrality classes.  Also given are the systematic
  uncertainties resulting from the variation of the Glauber model
  parameters as described in the text.  The first corresponds to a
  centrality selection in the impact parameter, while in the case of
  the other two the fraction of the total cross section is selected
  from the observables \nhitsS\ or \ntracks\ and thus also takes into
  account their different sensitivities in the centrality selection.}
\label{tab:centralities}
\end{table*}
%

\subsection{Systematic uncertainties}
\label{sect:syst_error}

The systematic uncertainties on the \npartav\ determination can be
separated into two categories.  On one side there are those related to
the input parameters of the Glauber MC model itself, on the other side
the different experimental centrality estimators introduce systematic
deviations from the ideal model scenario.

To investigate systematic effects due to the model parameters, the
fits were repeated with Glauber MC simulations based on different
input parameters.  The inelastic nucleon-nucleon cross section was
varied up to $\sinel^{\rb{NN}} = 30$~mb, which causes a change of the
extracted \npartav\ values of around 5\%, almost independent of
centrality.  We also tested the impact of different parameters for the
radial charge density distribution: the values $R = 6.75$~fm and $a =
0.623$~fm, as well as $R = 6.35$~fm and $a = 0.423$~fm, were used,
as motivated by \cite{Hahn:1956zz} and \cite{Buss:2011mx}.  In
comparison to the default parameter selection this causes a variation
of \npartav\ of at most 15\% for very peripheral collisions, which
decreases to 3-4\% for very central ones.  Furthermore, the fits
were repeated with an inter-nucleon exclusion distance of
$d_{\rb{min}} = 0$.  It was found that this modification affects the
resulting \npartav\ by not more than 3\%.  The final systematic
uncertainty of \npartav\ from the Glauber MC model is determined as
the maximal deviation between the different parameter sets and
procedures to the default version for each centrality.

%
\begin{figure}
\begin{center}
\includegraphics[width=0.9\linewidth]{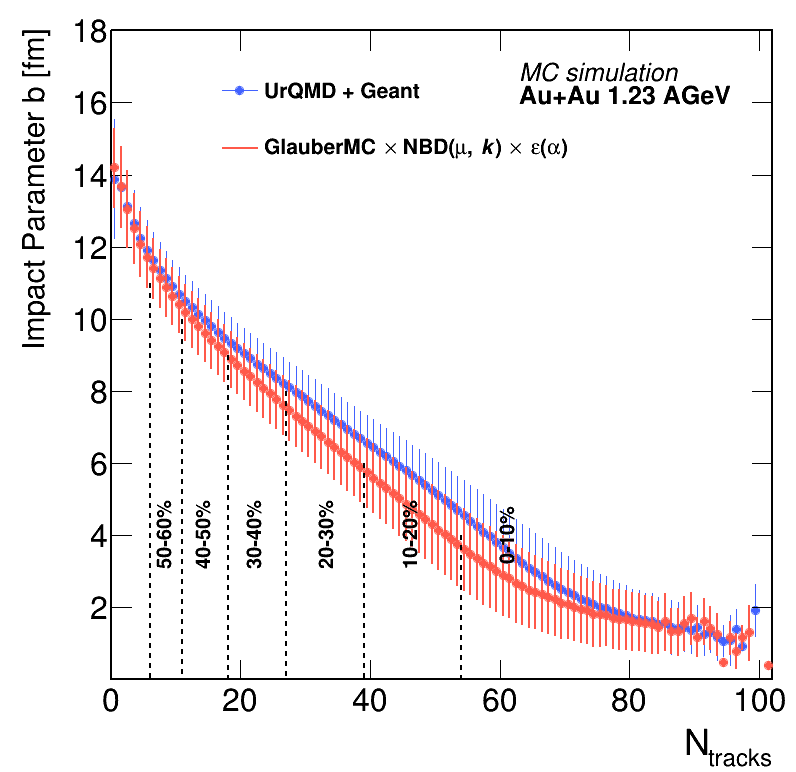}
\includegraphics[width=0.9\linewidth]{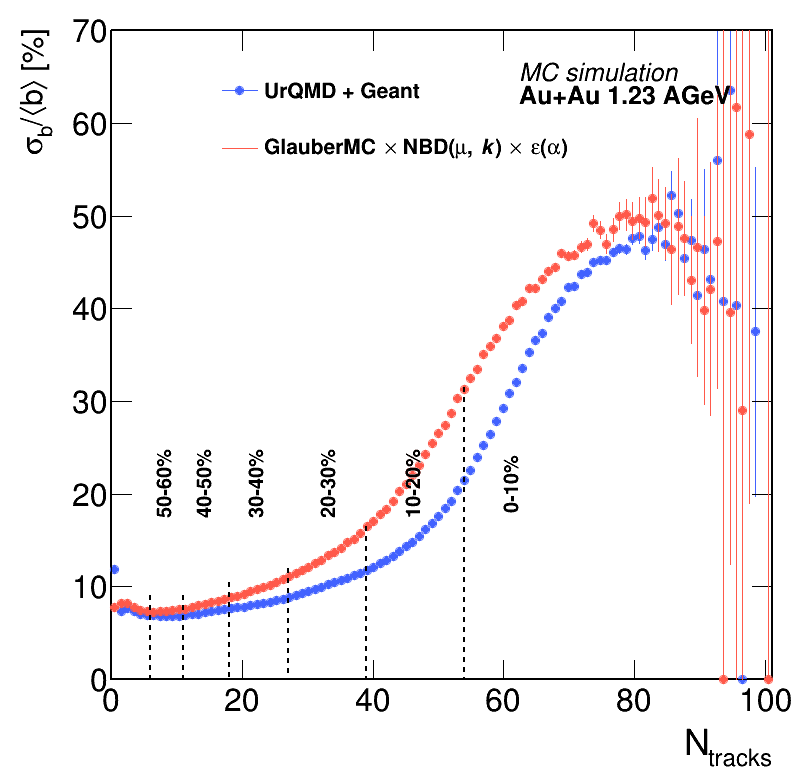}
\end{center}
\caption{The same as in fig.~\ref{fig:nhits_resolution}, but for 
  the total number of tracks \ntracks\ as obtained from UrQMD events,
  filtered through the detector simulation (blue symbols), and as
  taken from the Glauber MC model fit (red symbols).}
\label{fig:ntracks_resolution}
\end{figure}
%

While the above discussed systematic effects are inherent of the
ideal geometrical Glauber MC model implementation, additional effects
have to be taken into account for the experimental centrality
estimators.  On one side, these are due to different levels of
background and distortions, which cause deviations from the ideal
multiplicity estimation.  On the other side, the estimators provide
different resolutions for the event centrality due to the
reconstructed event-wise multiplicities (\ntracks\ is on average
smaller than \nhitsS, because of the more stringent cuts).  Therefore,
\npartav\ was calculated again according to the corresponding fraction
of the total cross section as determined by \nhitsS, respectively
\ntracks, instead of using cuts on the impact parameter.  It is found
that the resulting value of \npartav\ differs by less than 1\% in the
case of \nhitsS\ from the value given in table~\ref{tab:centralities}.
For \ntracks\, a slightly larger deviation of 3\% is observed for the
10\% most central events, while it is 1\% in the other centrality
ranges.

\subsection{Centrality classes}
\label{sect:cent-class}

The above described Glauber MC model implementation results in a total
cross section for \AuAu\ of
$\sigma_{\rb{tot}}^{\rb{Au+Au}} = (6833 \pm 430)$~mb.  The events can thus
be categorised into centrality classes according to their fraction of 
$\sigma_{\rb{tot}}^{\rb{Au+Au}}$.  Using the fits shown in
figs.~\ref{fig:ntracks_data} and \ref{fig:nhits_data_total} this
can be translated into intervals of \ntracks, respectively \nhitsS,
corresponding to these centrality classes.
Figure~\ref{fig:nhits_correlation_urqmd} shows the anti-correlation of
\nhitsS\ and the impact parameter $b$, obtained from the Glauber MC
model (upper panel) and from simulated UrQMD events (lower panel).
Over a large part of the total cross section, a well defined
anti-correlation is visible in both, allowing a good definition of the
different centrality classes, here indicated by the dashed lines.  As
illustrated in the upper panel of fig.~\ref{fig:nhits_resolution},
these two anti-correlations are in agreement well within their errors.
This demonstrates that a meaningful comparison of centrality selected
events is possible between data and model using \nhitsS.  From these
anti-correlations also the resolution in the impact parameter can be
derived (see upper panel of fig.~\ref{fig:nhits_resolution}).  It is
here defined as the ratio of the dispersion $\sigma$ obtained with a
Gaussian fit to the mean of the corresponding $b$ distribution at a
given value of \nhitsS.  Also here, a reasonable agreement is achieved
between UrQMD and the Glauber MC model fit.  The resolution is found
to be below 15\% for all centralities, except for the very central
interval 0-5\%.  Figure~\ref{fig:ntracks_resolution} shows the
same comparison for the observable \ntracks.

%
\begin{figure}
\begin{center}
\includegraphics[width=0.9\linewidth]{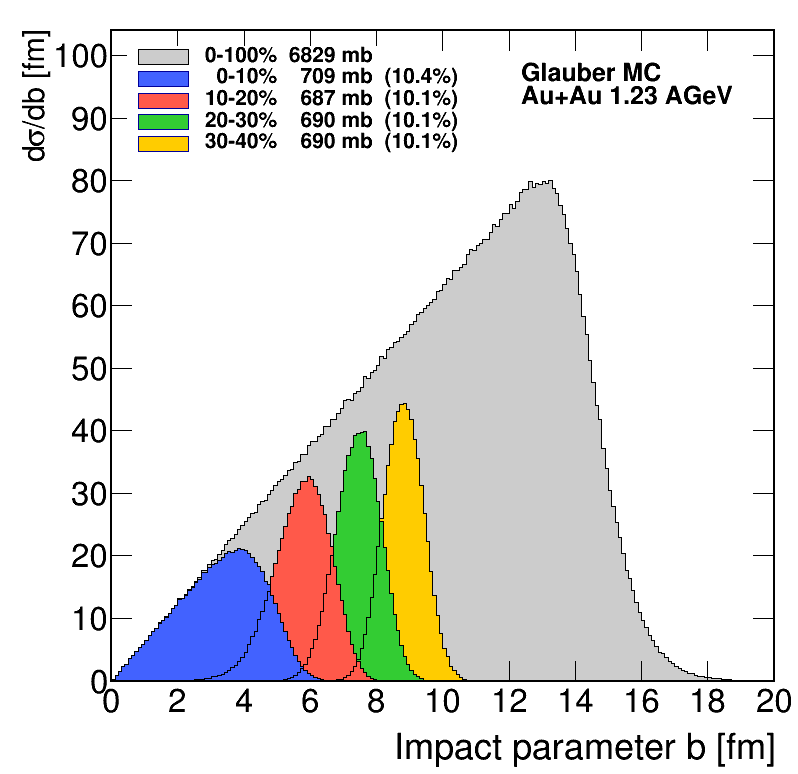}
\includegraphics[width=0.9\linewidth]{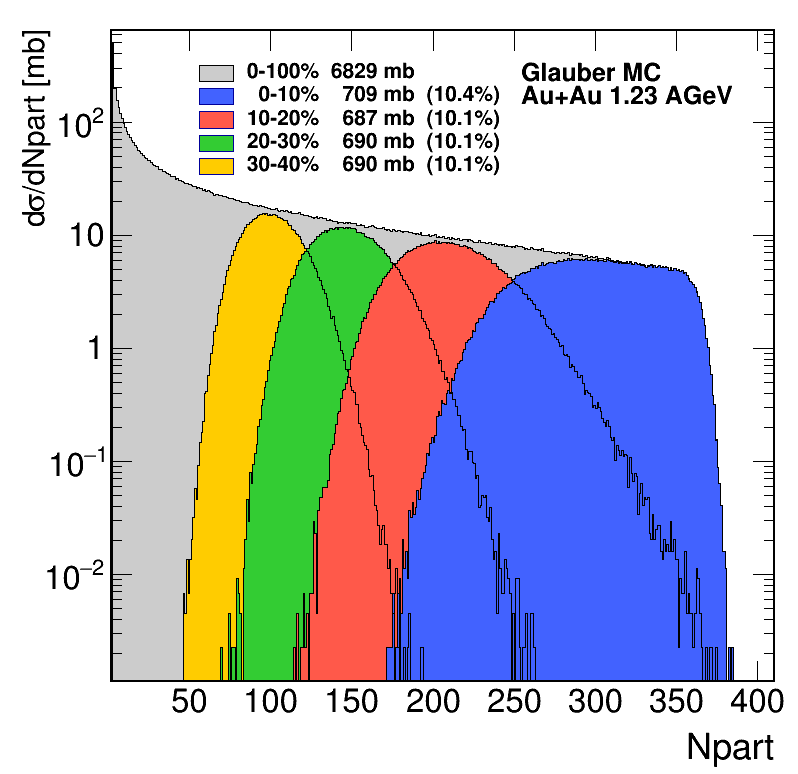}
\end{center}
\caption{Upper panel: The distributions of the impact parameter $b$
  calculated with the Glauber MC model.  The colored distributions
  represent the four most central centrality classes selected by the
  number of hits in the TOF and RPC detectors \nhitsS.
  Lower panel: The corresponding distributions of the number of
  participants \npart.}
\label{fig:npart_strk}
\end{figure}
%

The distributions of the impact parameter $b$ and number of
participants \npart\ for different centrality classes, as selected by
\nhitsS, are shown in fig.~\ref{fig:npart_strk}.  The averaged values
are summarized in table~\ref{tab:centralities}.  Even though the
distributions have some overlap (the corresponding RMS values can also
be found in table~\ref{tab:centralities}), a clear separation of the
centrality classes is possible.  
%

%
\section{Conclusions}
\label{sect:conclution}

The Glauber Monte Carlo model has been employed for the event
characterization of \AuAu, as measured with HADES at GSI-SIS18.  The
different experimental centrality estimators, number of reconstructed
tracks, \ntracks, and number of recorded hits in TOF and RPC, \nhitsS,
can successfully be described by the model fits.  Based on these fits
the events can thus be categorised in different centrality classes
with corresponding average number of participating nucleons \npartav\
and impact parameter \bav.  This procedure provides a well defined and
universal approach to determine the centrality dependences of many
observables measured by HADES, such as strange hadron production
\cite{Adamczewski-Musch:2017rtf}, dileptons and flow patterns.

%

\begin{acknowledgement}
The HADES Collaboration gratefully acknowledges the support by
PTDC/FIS/113339/2009 LIP Coimbra, NCN grant 2013/10/M/ST2/00042 SIP JUC
Cracow, Helmholtz Alliance HA216/EMMI GSI Darmstadt, VH-NG-823, Helmholtz
Alliance HA216/EMMI TU Darmstadt, 283286, 05P12CRGHE HZDR Dresden,
Helmholtz Alliance HA216/EMMI, HIC for FAIR (LOEWE), GSI F$\&$E
Goethe-Universit\"{a}t, Frankfurt VH-NG-330, BMBF 06MT7180 TU M\"{u}nchen,
Garching BMBF:05P12RGGHM JLU Giessen, Giessen UCY/3411-23100, University
Cyprus, CNRS/IN2P3 IPN Orsay, Orsay, MEYS LM2015049,
CZ.02.1.01/0.0/0.0/16\_013/0001677 NPI CAS Rez.
\end{acknowledgement}

%

%

%
\end{document}